\documentclass[aps,prl, twocolumn]{revtex4-1}
\usepackage{amsfonts}
\usepackage{amsmath}
\usepackage{amssymb}
\usepackage{graphicx}
\usepackage{times}
\usepackage{physics}
\usepackage[colorlinks=true,linkcolor=blue,citecolor=red,plainpages=false,pdfpagelabels]
{hyperref}
\usepackage[encapsulated]{CJK}

\usepackage{color}

\providecommand{\U}[1]{\protect\rule{.1in}{.1in}}

\begin{document}

\title{Direct parameter estimations from machine-learning enhanced quantum state tomography}
\author{Hsien-Yi Hsieh,$^{1}$ Jingyu Ning,$^{1}$ Yi-Ru Chen,$^{1}$ Hsun-Chung Wu,$^{1}$ Hua Li Chen,$^{2}$  Chien-Ming Wu,$^{1}$ and Ray-Kuang Lee$^{1,2,3,4}$}

\affiliation{$^{1}$Institute of Photonics Technologies, National Tsing Hua University, Hsinchu 30013, Taiwan\\
$^{2}$Department of Physics, National Tsing Hua University, Hsinchu 30013, Taiwan\\
$^{3}$Physics Division, National Center for Theoretical Sciences, Taipei 10617, Taiwan\\
$^{4}$Center for Quantum Technology, Hsinchu 30013, Taiwan}
 \email{rklee@ee.nthu.edu.tw}

\date{\today}
\begin{abstract}
With the capability to find the best fit to arbitrarily complicated data patterns, machine-learning (ML) enhanced quantum state tomography (QST) has demonstrated its advantages in  extracting  complete information about the quantum states.
Instead of using the reconstruction model in training a truncated density matrix, we develop a high-performance, lightweight, and easy-to-install supervised characteristic model by generating  the target parameters directly.
Such a characteristic model-based ML-QST can avoid the problem of dealing with large Hilbert space, but keep feature extractions with high precision. 
With the experimentally measured data generated from the balanced homodyne detectors, we compare the degradation information about  quantum noise squeezed states predicted by the reconstruction and characteristic models, both give agreement to the empirically fitting curves obtained from the covariance method.
Such a ML-QST with direct parameter estimations illustrates a crucial diagnostic toolbox for  applications with squeezed states, including  advanced gravitational wave detectors, quantum metrology, macroscopic quantum state generation, and quantum information process.
\end{abstract}

\maketitle

\section{Introduction}
Due to  unavoidable coupling from the noisy environment,  we need to have the ability to fully and precisely characterize the quantum features in a large Hilbert space.
In general,  the reconstruction is not on the quantum state, but the corresponding density matrix as the degradation makes the target quantum state into a mixed state. 
For continuous variables with infinite dimensions,  by utilizing multiple phase-sensitive measurements through homodyne detectors, quantum state tomography (QST) has provided us with a useful  tool for reconstructing quantum states~\cite{Hradil, QST}.
Nowadays, in a variety of  quantum systems, including quantum optics~\cite{QST-book,QST-Furusawa}, ultracold atoms~\cite{QST-atom1, QST-atom2}, ions~\cite{QST-ion-1, QST-ion-2}, and  superconducting circuit-QED devices~\cite{QST-SC}, QST has been successfully implemented as a crucial diagnostic toolbox for quantum information process.

By estimating the closest probability distribution to the data for any arbitrary quantum states, the maximum likelihood estimation (MLE) method is one of the most popular methods in reconstructing quantum states ~\cite{MLE}. 
However, MLE suffers from the overestimation problem, as the required amount of measurements to reconstruct the quantum state in multiple bases  increases exponentially with the number of involved modes. 
To overcome the overestimation in MLE, several alternative algorithms are proposed by assuming some physical restrictions imposed upon the state in question, such as permutationally invariant tomography~\cite{permu}, quantum compressed sensing~\cite{compress}, tensor networks~\cite{tensor-1, tensor-2}, generative models~\cite{generative}, and restricted Boltzmann machine~\cite{RBM}. 
Instead, with the capability  to find the best fit to arbitrarily complicated data patterns with a limited number of parameters available,   machine-learning (ML) enhanced QST was implemented experimentally, demonstrating  a fast, robust, and precise QST for continuous variables~\cite{RBM, QML-1, QML-2, PRL-22}.

However, in dealing with  continuous variables,  even truncating the Hilbert space into a finite dimension, a very large amount of data are still needed in reconstructing a truncated density matrix.
In this work, instead of training the machine on the reconstruction model, alternatively, we develop a characteristic model-based ML-QST by skipping the training on the truncated density matrix.
Such a characteristic model-based ML-QST can avoid the problem of dealing with large Hilbert space but keep feature extraction with high precision. 
With the prior knowledge of the experimentally measured data generated from the balanced homodyne detectors, the direct parameter estimations,  including the average photon numbers in the pure squeezed states, squeezed thermal states, and thermal reservoirs, agree with those acquired from the reconstruction model.
Compared to the empirically fitting curves obtained from the covariance matrix, our characteristic model-based ML-QST also reveals all  the degradation information about  quantum noise squeezed states, indicating the loss and phase noises in the measured  anti-squeezing.
With the ability to instantly monitor quantum states, as well as to make feedback control possible, our experimental implementations  illustrate a crucial diagnostic toolbox for the applications with squeezed states.
Based on the direct parameter estimations from this ML-QST, applications to the advanced gravitational wave detectors, quantum metrology, macroscopic quantum state generation, and quantum information process can be readily realized.

The paper is organized as follows: in Section II, we introduce the supervised machine-learning enhanced quantum state tomography based on the convolutional neural network (CNN).  
Then, the implementations of the reconstruction model  and characteristic model are illustrated in Section II (a) and II (b), respectively. 
The comparisons on the predicted average photon numbers, as well as the squeezing-anti-squeezing curve to the experimental fittings, are demonstrated in Section III, validating the feature extraction from our direct parameter estimations.
Finally, we summarize this work with some perspectives in Conclusion.

\begin{figure}[h]
\includegraphics[width=8.4cm]{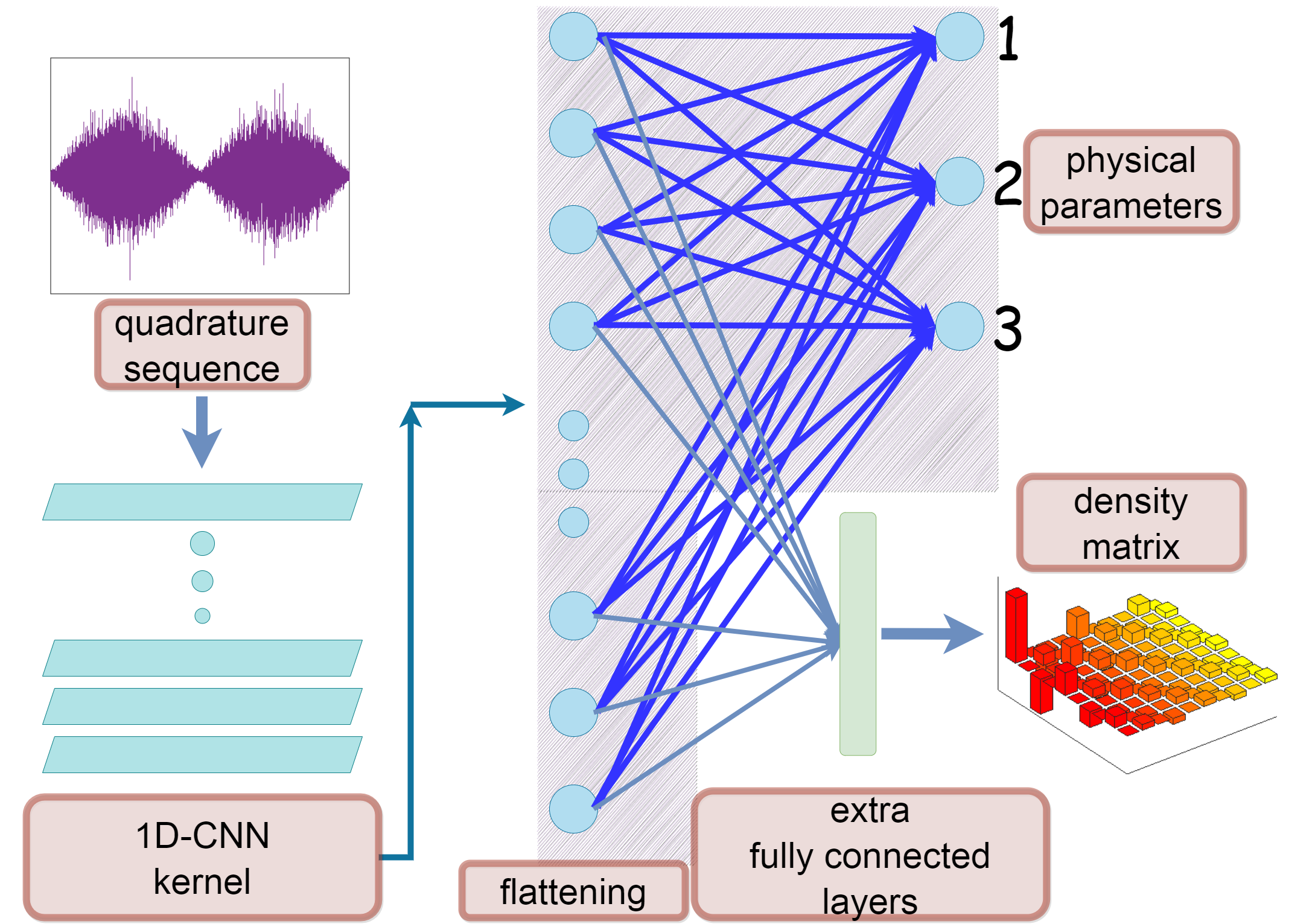}
\caption{Demonstration of direct parameter estimations with machine learning.  Here, the noisy data of quadrature sequence obtained by quantum homodyne tomography in a single-scan of LO phase from $0$ to $2\pi$ are fed to the convolutional layers, denoted as 1D-CNN kernel. Then, after flattening, either the density matrix is reconstructed through extra fully connected layers (the reconstructed model), or the physical parameters are predicted directly (the characteristic model, marked with the shadowed background).
The function of neurons marked as $(1, 2, 3)$ denotes inferring the value of $(r, \theta, n_{th})$, directly.}
\end{figure}

\section{Supervised machine-learning enhanced quantum state tomography}

When applying MLE to reconstruct the target quantum state, the data acquisition is performed by the balanced homodyne detectors based on the  covariance method or nullifiers~\cite{witness, comb, nullifier}.
However,  at least three measurements must be performed at a fixed (but different) local oscillator (LO) phase in order to estimate the probability distribution function in different quadratures. 
To reduce unwanted uncertainty in the fixed quadrature, therefore,  a precise phase locking for the LO phase is also needed.
However, in the homodyne experiments, the repeatability of the PZT drifts owing to the airflow and the temperature difference, resulting in  introducing additional (phase) noises into the measuring system. 
Moreover, the validation of this method relies on the Gaussian properties of reconstructed states~\cite{8}. Nevertheless,  information in different LO phases is missing due to the selected measurements only at three LO phases.

Instead, by scanning the LO phase from $0$ to $2\pi$, referring as a single-scan of quadrature sequence data, $X_{\theta}$, our homodyne measurements contain all the information at different LO phases~\cite{PRA}.
Intrinsically, the phase noise automatically is counted in our ML-QST~\cite{PRL-22}. 
A fast QST is possible with such a single-scan measurement by just varying the LO phase.
Here, the quadrature sequence data  $X_{\theta}$ shares the similarity to the voice (sound) pattern in a time series~\cite{7}.
With the prior knowledge of squeezed states, a supervised ML with CNN configuration is introduced in this work.

As illustrated in Fig. 1, by feeding  noisy data of quadrature sequence acquired by quantum homodyne tomography into $17$ convolutional layers, we take  advantage of  good generalizability in applying CNN ~\cite{generalizability}.
In our one-dimensional (1D)-CNN kernel, there are $5$ convolution blocks used, each of which contains $2$ convolution layers (filters) in different sizes.
In order to tackle the gradient vanishing problem, which  commonly happens in the deep CNN when the number of convolution layers increases, some shortcuts are also introduced among the convolution blocks~\cite{GV}.
Nevertheless, after flattening the 1D-CNN kernel, we either  apply  extra fully connected layers to reconstruct the truncated density matrix (coined as the reconstructed model), or predict physical parameters directly (coined as the characteristic model).
Below, the details and differences in the reconstruction model and characteristic modes are described.

\subsection{Reconstruction model:}
The target of implementing the reconstruction model is to predict the truncated density matrix. 
In the quantum noise squeezing experiments,  we have three families of possible states, i.e., pure squeezed state $\rho^{sq}$, squeezed thermal states $\rho^{sq}_{th}$, and thermal states $\rho_{th}$~\cite{sq-thermal-1, sq-thermal-2, sq-thermal-3, 5}.
These three families can be described uniformly by a generic formula for squeezed thermal states, i.e.,  
\begin{eqnarray}
\hat{\rho} = \hat{S}(r,\theta) \hat{\rho}_{th}(n_{th})\hat{S}^{\dag}(r,\theta).
\end{eqnarray}
As shown in Eq. (1), we have three characteristic parameters, $r$, $\theta$, and  $n_{th}$, corresponding to the squeezing ratio, squeezing angle, and the average photon number, respectively. 
Here, $\hat{S} (r, \theta) = \text{exp}[\frac{1}{2} ( \xi^\ast \hat{a}^2 - \xi \hat{a}^{\dag 2})]$ denotes the 
squeezing transformation, with $\xi \equiv r\, \text{exp}(i\, \theta)$; $r \in [0, \infty]$ and $\theta \in [0, 2\pi]$.

One can see that when $r = 0$, Eq. (1) describes the thermal states with the average photon number $n_{th}$, reflecting the corresponding temperature in the thermal reservoir, i.e., $  \bar{n}^{-1} =  \text{exp}[\hbar \omega/k_B T]-1$.
 However, when $n_{th} = 0$, Eq. (1) gives the pure squeezed vacuum state, characterized by its squeezing ratio $r$ and the squeezing angle $\theta$. 
In training the machine, a uniform sampling with different physical parameters $(r, \theta, n_{th})$ is applied for generating  the simulated quadrature sequence.

The task of  our reconstruction model can be formulated as mapping the estimated function to a truncated density matrix, i.e.,  $f_{\text{est}}: X_{\theta}\rightarrow \hat{\rho}_{m \times m}$. 
Here,  $m$ denotes the dimension of our truncated Hilbert space in the number state basis.
To avoid non-physical states, we impose the positive semi-definite constraint into the predicted density matrix.
An auxiliary (lower triangular) matrix is introduced before generating the predicted factorized density matrix  through the Cholesky decomposition, i.e., $\hat{\rho}_{m \times m} \equiv L_{m \times m}\, L_{m \times m}^{\ast}$.
The training set for the quadrature data $\{X_\theta^j\}$  is the set formed by  
\begin{eqnarray}
\left \{X_\theta^{j}, L_{m \times m}^{j} \vert \text{dim}(X^j_\theta) = 4096, \theta \in [0,2 \pi], j = 1,2,3 \dots N\right \}.
\end{eqnarray}
Here, $N$ is the number of the training set, and  $\text{dim}(X^j_\theta) = 4096$ is chosen for the number of sampling data in a quadrature sequence.
Our target is  training the machine to  learn the function $f_{\text{est}}$, which can be mapped from $X_{\theta}$  to $L_{m \times m}$.
This estimation function can be approximated by a deep neural network which is parametrized by trainable weight variables $W^{l}$, with $l$ corresponding to the $l$-th layer in the deep neural network, i.e., 
$f_{\text{est}} \sim f^{l}(\dots f^{2}(f^{1}(X_{\theta}, W^{1}), W^{2}) \dots W^{l})$.
The training process is to minimize the mean squared error (MSE); while the optimizer used for training is Adam. 
We take the batch size as $32$ in the training process. By this setting, the network is trained with $70$ epochs to decrease the loss (MSE) up to $5\times 10^{-6}$.
Moreover,  the normalization is also applied during  the training process in order to  ensure that the trace of the output density matrix is kept as $1$. 
Furthermore, to improve the performance in feature extraction and to reduce the number of parameters, the dense connection is also introduced in our 1D-CNN kernel~\cite{4, 12}. This makes the our 1D-CNN model more efficient and lightweight.
Finally, as the schematic shown in Fig. 1, after flattening, the predicted matrices are inverted to reconstruct the density matrices in truncation, 

\begin{table}
\begin{tabular}[t]{|c|r|}
\hline
Layer name	& Parameters\\
\hline
Conv\_1d\_layer1 	& [4, 96]\\
\hline
Conv\_1d\_block\_a  &  [4, 96]\\
&  [4, 96] \\
\hline
Transition Layer 1: Conv\_1d	& [1, 48] (stride = 4)\\
\hline
Conv\_1d\_block\_b1 & 	[4, 64]\\
& [4, 64]\\
\hline
Conv\_1d\_block\_b2 &	[4, 64]\\
&[4, 64]\\
\hline
Transition Layer 2: Conv\_1d&	[1, 64] (stride = 4)\\
\hline
Conv\_1d\_block\_c1 &	[4, 128]\\
&[4, 128]\\
\hline
Conv\_1d\_block\_c2 &[4, 128]\\
&[4, 128]\\
\hline
Transition Layer 3: Conv\_1d &	[1, 96] (stride = 4)\\

\hline
Conv\_1d\_layer4 &	[4, 96] (stride = 2)\\
\hline
Conv\_1d\_layer5 &	[2, 128] (stride = 2)\\
\hline
Conv\_1d\_layer6 &	[2, 48] (stride = 2)\\
\hline
\end{tabular}
\caption{Hyper-parameters (filter sizes of each layer) used in our 1D CNN.}
\end{table}

By considering our quantum optics experiments with the maximum squeezing level up to $10$~dB and 
the maximum anti-squeezing level up to $20$~dB, we keep the sum in the probability up to $0.9999$ by truncating the photon number to $m = 35$.
More than one million data sets (exactly, $N= 1,200,000$) are fed into our machine with a variety of pure squeezed states, squeezed thermal states,  and thermal states in different squeezing levels,  quadrature angles, and reservoir temperatures.
All the training is carried out with the Python package {\it tensorflow.keras} performed in GPU  (Nvidia Titan RTX).
When well-trained (typically in less than one hour), the execution time for our machine-learning enhanced QST takes the average cost time $38.1$ milliseconds (by averaging $100$ times) in a standard GPU server.

Regarding the hyper-parameters (filter sizes of each layer), in Table 1, we provide information about the architecture and parameters used in our 1D-CNN kernel. 
The parameters in this table correspond to the kernel size and channel length. For example, $[4,96]$ means that kernel size $=4$ and channel length $=96$. Here, the size of our density matrix  is $35 \times 35$.  There are also convolutions in shortcuts with dense connections~\cite{12}.

\subsection{Characteristic model:}

In general, the supervised ML is performing a regression task, predicting a truncated density matrix for the quantum state tomography. 
However, as shown in Eq. (1), the target mixed state is just a linear combination of three families composed of pure squeezed states, squeezed thermal states, and thermal states. 
These physical states can be  basically described by a few simple physical parameters. 
In addition to reconstructing the density matrix, one can also train a machine to predict parameters directly, coined as a {\it characteristic model}. 

In the quantum noise squeezing experiments, the parameter set defined by  $(r, \theta, n_{th})$ should provide enough information in the output measurements, which are the measured squeezing level (SQZ) and the anti-squeezing level (ASQZ). 
This characteristic model can help us to  avoid the problem that occurs in dealing with high-dimensional Hilbert space. 
Compared to the reconstruction model, now the task of our supervised estimation is mapping the estimated function to the physical parameters directly, i.e., $f_{\text{est}}: X_{\theta}\rightarrow (r, \theta, n_{th})$.

As marked in Fig. 1 with the shadowed background, we can directly generate these three physical parameters, without bothering additional  extra fully connected layers.
In this characteristic model, after the convolution kernel completes the feature extraction, we do not need to apply the fully connected layers, but just perform a linear transformation to predict the characteristic values of  quantum states. 
In addition, in the characteristic model, we take the batch size as $32$ in the training process. By this setting, the network is trained with $30$ epochs to guarantee the error (MSE) no larger than $0.03$.

The advantages of applying this characteristic model come from the absence in dealing with any post-processing.
Of course, one can calculate these physical parameters with  help of the reconstructed density matrix.
However, as fewer model parameters (architecture size) are involved, we also avoid the possible errors caused due to the truncation in the density matrix.  In the following, we will demonstrate the implementation of this characteristic model-based ML in the laboratory, by directly and quickly inferring the value of $(r, \theta, n_{th})$ in the quantum noise squeezing experiments.

\begin{figure}[t]
\includegraphics[width=8.4cm]{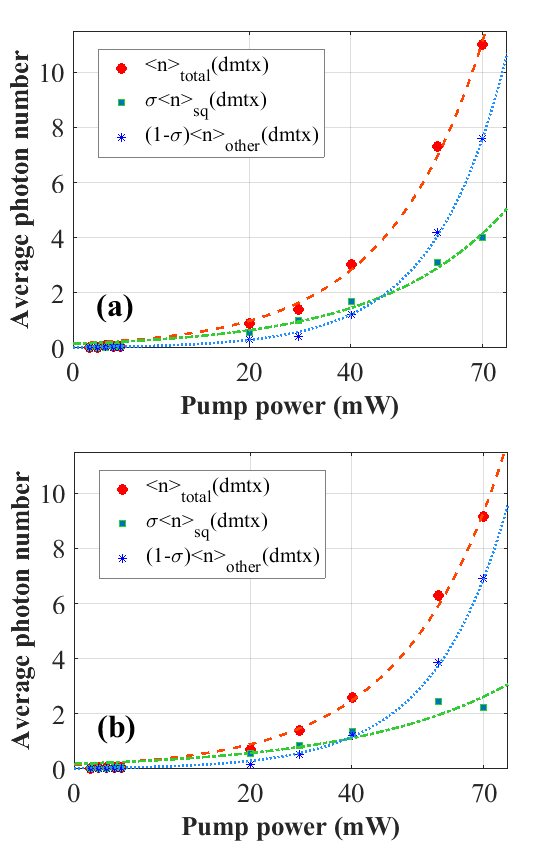}
\caption{The comparisons on the predicted average photon number, as a function of pump power, between (a) characteristic model and (b) reconstruction model. 
In the characteristic model, we directly generate the parameter estimations (para est) for the average photon number in the measured data $\langle n\rangle_{\text{total}}$ in Red, the pure squeezed state $\langle n\rangle_{\text{sq}}$ in Green, and non-pure components $\langle n\rangle_{\text{other}}$ in Blue. 
However, in the reconstruction model, we need to apply the singular value decomposition to the predicted density matrix (dmtx) first, revealing the coefficient $\sigma$ for the pure squeezed state and $(1-\sigma)$ for non-pure components. 
}
\end{figure}

\section{Comparison between the reconstruction and characteristic models}

As the report in Ref.~\cite{PRL-22}, our quantum noise squeezed states are generated through a bow-tie optical parametric oscillator cavity with a periodically poled KTiOPO$_4$ (PPKTP) inside, operated below the threshold at the wavelength of $1064$ nm. 
Experimentally, the quantum homodyne tomography is performed by collecting quadrature sequence with the spectrum analyzer  at $2.5$ MHz with $100001$ data points, $100$ kHz RBW (resolution bandwidth), and $100$ Hz  VBW (video bandwidth).
The phase of LO is scanned with a $1$ Hz triangle wavefunction.
While the pump power increases to $70$ mW,  the measured noise levels for squeezing (SQZ) and anti-squeezing (ASQZ) in decibel (dB) are $8.37$ and $17.00$, respectively.
In training the reconstruction model, we use a ``uniform distribution" to sample the value of LO angle, with $4096$ sampling  points fed from the experimental datasets ($5,000,000$ data points).
Our well-trained reconstruction model-based ML-QST has demonstrated its advantage in keeping the fidelity in the predicted  density matrix as high as  $0.99$~\cite{PRL-22}.

Now, to verify the physical parameter estimation with the characteristic model, in Fig. 2, we compare the predicted average photon number, as a function of pump power, between (a) characteristic model and (b) reconstruction model. 
 As the pump power increases,  both the characteristic and reconstruction models give great agreement in predicting the three curves of average photon numbers for the measured data $\langle n\rangle_{\text{total}}$, the pure squeezed state $\langle n\rangle_{\text{sq}}$, and non-pure components $\langle n\rangle_{\text{other}}$, denoted as (para est) for the parameter estimation and (dmtx) for the density matrix in Figs. 2(a) and 2(b), respectively.
Not only the tendency of monochromatic increment in these three curves, both models also reveal the cross-over between the pure squeezed states and non-pure components, as shown in the Blue- and Green-colors.
This cross-over indicates that the non-pure components become  dominant parts at a higher pump power, which makes the quantum noises degraded, resulting in ASQZ being larger than SQZ.

We want to remark that unlike the reconstruction model, the physical parameters are predicted directly from the characteristic model without any post-data processing. 
 However, in the reconstruction model, we need to apply the singular value decomposition to the predicted density matrix (dmtx) first. Then, only with the obtained coefficient $\sigma$ for the pure squeezed state, the weighting in the non-pure components $(1-\sigma)$ can be known. 

\begin{figure}[t]
\includegraphics[width=8.4cm]{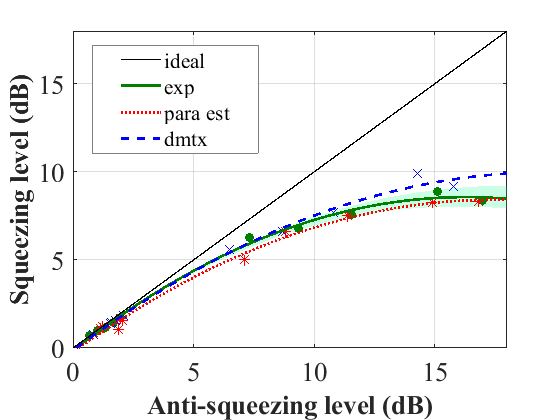}
\caption{Degradation in squeezed states, i.e., Squeezing level versus Anti-squeezing level.  Ideally, the squeezing and anti-squeezing levels should locate along the Black line (ideal). However, as shown with the typical experimental data, marked in Green dots, there exists a discrepancy between the measured squeezing and anti-squeezing levels. By taking the loss and phase noise into account, based on Eqs. (3-4), the optimal fitting curve is depicted in Green-color (exp), with the corresponding standard deviation shown by the shadowed region. Moreover, our ML-QST based on the reconstruction model (dmtx) and characteristic model (para est) both give agreement to experimental data, depicted in Blue-dashed and Red-dotted curves, respectively.}
\end{figure}

As  a crucial diagnostic toolbox for practical applications, we also compare our ML-QST, both on the reconstruction and characteristic models, with the experimental fitting curves on the degradation in squeezed states.
In experiments, the degradation in quantum noise squeezing is typically described by the squeezing versus anti-squeezing curve, as shown in Fig. 3.

Ideally, without any degradation, the squeezing and anti-squeezing levels should be the same, located along the Black line in Fig. 3.
However, the phase noise and loss mechanisms coupled with the environment and surrounding vacuum set the limit on the measured squeezing level.
Empirically, to estimate the loss and phase noises, not a single set of quadrature data but a series of sets of quadrature data must be performed, in order to have accurate fitting parameters for exp-fitting (co-variance fitting). 
By taking the optical loss (denoted as $L$) and phase noise (denoted as $\theta$) into account, the measured squeezing $V^{\text{SQZ}}$ and anti-squeezing $V^{\text{ASQZ}}$ levels can be modeled as
\begin{eqnarray}
&&V^{\text{SQZ}} = (1-L) [V^{\text{SQZ}}_{id}\times \cos^2 \theta + V^{\text{ASQZ}}_{id}\times \sin^2\theta]+L,\\
&&V^{\text{ASQZ}} = (1-L)[V^{\text{ASQZ}}_{id}\times \cos^2\theta + V^{\text{SQZ}}_{id}\times \sin^2\theta]+L,
\end{eqnarray}
where $V^{\text{SQZ}}_{id}$ and $V^{\text{ASQZ}}_{id}$ are the squeezing and anti-squeezing levels in the ideal case.
In Fig. 3, we also show the optimal fitting curve obtained by the orthogonal distance regression  in Green-color, with the corresponding  standard deviation (one-sigma variance) shown by the shadowed region.
As we show in Fig. 3, an accurate EXP-fitting can only be obtained by performing many (in our illustration, 12) different pump power levels.

Moreover, the success of EXP-fitting relies on the common belief that as long as the system is stable, the loss and phase noises can be estimated.
Nevertheless, as we illustrated, such a common belief is only valid at a low degree of squeezing (less than 5 dB). On the contrary, when the pump power increases, many additional effects, such as the heating in crystals, shift of resonance frequency, and/or other nonlinear mechanisms, occur, resulting in the increment in loss~\cite{PRL-22}.

On the contrary, only with a  single-scan measurement, our ML-QST based on the reconstruction model (dmtx) and characteristic model (para est) both give agreement to experimental data, depicted in Blue-dashed and Red-dotted curves, respectively.
The curves shown in Fig. 3 clearly demonstrate that  our well-trained ML-QST can  extract the degradation information in quantum states not only very precise, but also very fast. 
Compared to the time-consuming MLE, this feature paves the road toward a real-time and online QST~\cite{2, 4}. Application of this machine-learning enhanced QST has also been applied to the reconstruction of Wigner current~\cite{4}, which definitely can only be achieved with this methodology.

\section{Conclusion}
In summary, we develop a characteristic model to directly predict physical parameters in 1D-CNN configuration, without dealing with density matrix in a higher dimensional Hilbert space.
Based on the prior knowledge about target quantum states, the predicted physical parameters obtained by our  characteristic model are as good as those generated by a reconstruction model.
Through the validation with the experimentally measured data acquired from the balanced homodyne detectors,  agreement to the empirically fitting curves obtained from the covariance method is clearly demonstrated. 
Such a characteristic model-based ML-QST  can be easily installed on edge devices like FPGA as an in-line diagnostic toolbox for the applications with squeezed states, including the advanced gravitational wave detectors, quantum metrology, macroscopic quantum state generation, and quantum information process.

\section*{Acknowledgement}

This work is partially supported by the Ministry of Science and Technology, Taiwan under grants (MOST 110-2123-M-007-002, 110-2627-M-008-001),  the International Technology Center Indo-Pacific (ITC IPAC) and Army Research Office, under Contract No. FA5209-21-P-0158, and the Collaborative research program of the Institute for Cosmic Ray Research (ICRR), the University of Tokyo.

\end{document}